\documentstyle[11pt,fullpage,epic,eepic,epsfig]{article}
\input{epsf}

\newtheorem{definition}{Definition}[section]
\newtheorem{theorem}{Theorem}[section]

\newtheorem{coro}[theorem]{Corollary}
\newenvironment{proof}{\noindent{\bf Proof }}%
                {\hspace*{\fill}$\Box$\par\vspace{4mm}}

\newcommand{\remove}[1]{}

\begin{document}
\title{{\bf Consistent Data Checkpoints \\
in Distributed Database Systems: a Formal Approach}\thanks{This work was
done during a stay of Michel Raynal at the Dipartimento di Informatica e
Sistemistica of Rome  supported by a grant of  the {\em Consiglio Nazionale
delle  
Ricerche} in the context of the Short-Term-Mobility program.}}

\author{ 
Roberto {\sc Baldoni}, Francesco {\sc Quaglia}\\
Dipartimento  di Informatica e  Sistemistica\\   Universit\'{a} di Roma "La
    Sapienza"    \\    Via     Salaria       113,  Roma,   Italy         \\
    baldoni,quaglia@dis.uniroma1.it  
\and
Michel {\sc
    Raynal} \\ IRISA \\ Campus de Beaulieu \\ 35042 Rennes-Cedex,
    France. \\ raynal@irisa.fr 
}

\date{}

\maketitle

\begin{center}
Technical Report 22-97 \\
Dipartimento di Informatica e Sistemistica\\
Universit\'a di Roma "La Sapienza" \\
Luglio 1997 
\end{center}

\begin{abstract}
  
  Whether it is for  audit or for recovery  purposes, 
  data checkpointing is an important 
  problem of distributed database systems. Actually, transactions
  establish dependence relations on data checkpoints taken by data
  object managers. So, given an arbitrary set of data checkpoints
  (including at least a single data checkpoint from a data manager,  
  and at most a data
  checkpoint from each data manager), an important question is the
  following one: ``Can these data checkpoints be members of a same
  consistent global checkpoint?''.
  
  This paper answers this question by providing a necessary and
  sufficient condition suited for database systems. Moreover, to show
  the usefulness of this condition, two {\em non-intrusive} data
  checkpointing protocols are derived from this condition.  It is also
  interesting to note that this paper, by exhibiting 
  ``correspondences'', establishes a bridge between the data
  object/transaction model and the process/message-passing model.

\end{abstract}

%-------------------------------------------------------------------------
%-------------------------------------------------------------------------
\section{Introduction} 

Checkpointing the state of a database is important for audit or
recovery purposes.  When compared to its counterpart in 
distributed systems, the database
checkpointing problem has additionally to take into account the
serialization order of the transactions that manipulates the data
objects forming the database.  Actually, transactions create
dependencies among data objects which makes harder the problem of
defining {\em consistent} global checkpoints in  database systems.

Of course, it is always possible, in a database environment, to design
a special transaction, that reads all data objects and saves their
current values. The underlying concurrency control mechanism ensures
this transaction gets a consistent state of the data objects. But this
strategy is inefficient, intrusive (from the point of view of the
concurrency   control \cite{SON})  and not  practical   since,  a read only
transaction may take a very long time to execute  and may cause intolerable
delays for other transactions \cite{PK92}. Moreover, as pointed
out by Salem and Garcia-Molina \cite{SG-M87}, this strategy may drastically
increase the cost of rerunning aborted transactions. 
So, it is preferable to base global
checkpointing (1) on local checkpoints of data objects taken by their
managers, and (2) on a mechanism ensuring mutual consistency of local
checkpoints (this will ensure that it will always be possible to get 
consistent global checkpoints by piecing together local checkpoints).

In this paper we are interested in exploiting such an approach. We
consider a database in which each data object can be individually
checkpointed (note that a data object could include, practically, a
set of physical data items).  If these checkpoints are taken in an
independent way, there is a risk that no consistent global checkpoint
can ever be formed (this leads to the well known {\em domino effect}
\cite{R75}). So, some kind of coordination is necessary when local
checkpoints are taken in order they be mutually consistent. In this
paper we are interested in characterizing mutual consistency of local
checkpoints. More precisely, we are interested in the two following
points.
\begin{itemize}
\item First, we address the following question: ``Given an arbitrary
  set $S$ of checkpoints, can this set be extended to get a global
  checkpoint (i.e., a set including exactly one checkpoint from each
  data object) that is consistent?''. The answer to this question is
  well known when the set $S$ includes exactly one checkpoint per data
  object \cite{PK92}. It becomes far from being trivial, when the set $S$ is
  incomplete, i.e., when it includes checkpoints from only a subset of
  data objects. When $S$ includes a single data checkpoint, the
  previous question is equivalent to "Can this local checkpoint belong
  to a consistent global checkpoint?".
\item Then, we focus on data checkpointing protocols. Let us consider
  the property ``Local checkpoint $C$ belongs to a consistent global
  checkpoint''.  We  design  two non-intrusive   protocols. The  first  one
  ensures the   previous property  when  $C$ is  any local  checkpoint. The
  second one ensures it when $C$ belongs to a predefined set of local
  checkpoints.
\end{itemize}

The paper consists of 4 main sections. Section \ref{model} introduces
the database model we consider in this paper. Section \ref{CGC} defines
consistency of global checkpoints. Section \ref{DGC} answers the
previous question. To provide such an answer, it studies the kind of
dependencies both the transactions and their serialization order create
among checkpoints of distinct data objects. \remove{So, in this paper,
we follow the direction pointed out in \cite{BGRS91}, where it is said
that ``Although the problems of concurrency control and recoverability
are frequently discussed separately, they are actually closely
related''.} More specifically, it is shown that, while some data
checkpoint dependencies are causal, and consequently can be captured on
the fly \cite{L78}, some others are ``hidden'', in the sense that, they
cannot be revealed by causality.  It is the existence of those hidden
dependencies that actually makes non-trivial the answer to the previous
question.  Then, Section \ref{protocols} shows how the necessary and
sufficient condition stated in Section \ref{DGC}, can be used to design
``transaction-induced'' data checkpointing protocols ensuring the
property ``Local checkpoint $C$ belongs to a consistent global
checkpoint''. These protocols allow managers of data objects to take
checkpoints independently on each other\footnote{These checkpoints are
called {\em basic}. They can be taken, for example, during CPU idle
time.}, and use transactions as a means to diffuse information, among
data managers, encoding dependencies on the previous states of data
objects.  When a transaction that accessed a data object is committed,
the data manager of this object may be directed to take a checkpoint in
order previously taken checkpoints belong to consistent global
checkpoints.  Such a checkpoint is called {\em forced} checkpoint. This
is done by the data manager which exploits both its local control data
and the information exchanged at the transaction commit point. 

Last but not least, this paper can be seen as a bridge between the
area of distributed computing and the area of databases. For a long
time, databases have provided distributed computing with very
interesting problems and protocols related to data replication,
concurrency control, etc. We show here how database checkpointing can
benefit from studies that originated from distributed computing.
Actually, a similar question has been addressed in the context of
the  asynchronous process/message-passing  model  \cite{BHR,NX95}. 
In  this context a message 
establishes a simple relation between a pair of process local states. In the
database context, a transaction may  establish several relations
between states of data objects. So, albeit there are some
correspondences between the process/message-passing model and the data
object/transaction model\footnote{At some abstraction level, there are
similarities, on one side between processes and data objects, and on
the   other  side,   between  messages  and   transactions   (see  Section
\ref{distcomp}).}, it appears that 
extending process/message-passing model results to the context of
database transactions is not trivial as a transaction is "something"
more complicated than a message: a transaction is on several data
objects at a time, and accesses them by read and write operations
whose results depend on the serialization order.

%-------------------------------------------------------------------------
%-------------------------------------------------------------------------
\section{Database Model}
\label{model}

We consider a classical distributed database model. The system
consists of a finite set of data objects, a set of transactions and a
concurrency control mechanism (see \cite{BHG87,GR} for more details). 

\paragraph{Data objects.}
Each data object is managed by a data manager $DM$. A set of data
objects can be managed by the same data manager $DM$. For clarity, we
suppose that the set of data managed by the same $DM$ constitutes a
single logical data. So, there is a data manager $DM_x$ per data $x$.

\paragraph{Transactions.}
A transaction is defined as a partial order on {\em read} and {\em
write} operations on data objects and terminates with a {\em commit}
or an {\em abort} operation. $R_i(x)$ (resp. $W_i(x)$)
denotes a read (resp. write) operation issued by transaction $T_i$ on
data object $x$. Each transaction is  managed by an instance of the
transaction manager   (TM) that forwards its    operations to the scheduler
which   runs a specific  concurrency control  protocol. The  write set of a
transaction is the set of all the data objects it wrote. 

\paragraph{Concurrency control.}
A concurrency control protocol schedules read and write operations
issued by transactions in such a way that any execution of
transactions is {\em strict} and {\em serializable}.
This is not a restriction as concurrency control mechanisms used in
practice (e.g., two-phase locking 2PL and timestamp ordering) generate
schedules ensuring both properties \cite{BGRS91}. The {\it strictness}
property states that no data object may  be read or written until the
transaction  that currently writes it  either   commits or  aborts. So, a
transaction actually writes 
 a data object at its commit point. Hence, at some abstract level, which
is the one considered by our checkpointing mechanisms, transactions
execute atomically at their commit points. If a transaction is aborted,
strictness ensures no cascading aborts and the possibility to use {\em
before images} for implementing abort operations which restore the value
of an object before the transaction access.  For example, a 2PL
mechanism, that requires that transactions keep their write locks until
they commit (or abort), generates such a behavior \cite{BGRS91}. 

\paragraph{Distributed database.}
We  consider a distributed  database  as a  finite set of  sites, each site
containing  one   or    several (logical)  data objects.   So,  each   site
contains  one or more  data managers, and possibly an instance of the TM. 
TMs and DMs exchange messages on a communication network
which is asynchronous (message transmission times are unpredictable
but finite) and reliable (each message will eventually be delivered).

\paragraph{Execution.}
Let $T=\{T_1,T_2,\ldots,T_n\}$ be a set of transactions accessing a
set $D=\{d_1, d_2, \ldots ,d_m\}$ of data objects (to simplify
notations, data object $d_i$ is identified by its index $i$). An
execution $E$ over $T$ is a partial order on all read and write
operations of the transactions belonging to $T$; this partial order
respects the order defined in each transaction. Moreover, let $<_x$ be
the partial order defined on all operations accessing a data object
$x$,  i.e., $<_x$ orders all pairs of  conflicting operations (two operations
are  conflicting if they access  the  same object  and  one  of  them is  a
write). Given an 
execution $E$ defined over $T$,~~ $T$ is  structured as a {\em partial order}
$\widehat{T} =(T, <_T)$ where $<_T$ is the following (classical) relation
defined on $T$:

\[T_i<_TT_j \iff (i\neq j) \wedge (\exists x \Rightarrow (R_i(x) <_x
W_j(x)) \vee (W_i(x) <_x W_j(x)) \vee (W_i(x) <_x R_j(x)))\]

%-------------------------------------------------------------------------
%-------------------------------------------------------------------------
\section{Consistent Global Checkpoints}
\label{CGC}
%-------------------------------------------------------------------------
\subsection{Local States and Their Relations}
\label{localstates}
Each write on a data object $x$ issued by a transaction defines a new
version of $x$. Let $\sigma^{i}_{x}$ denote the $i$-th version of $x$;
$\sigma^{i}_{x}$ is called a {\em local state}.  Transactions
establish dependencies between local states. This can be formalized in
the following way.  When $T_k$ issues a write operation $W_k(x)$, it
moves the state of $x$ from $\sigma^{i}_{x}$ to $\sigma^{i+1}_{x}$.
More precisely, $\sigma^{i}_{x}$ and $\sigma^{i+1}_{x}$ are the local
states of $x$, just before and just after the
execution\footnote{Remind that, as we consider strict and serializable
  executions, ``Just before and just after the execution of $T_k$''
  means ``Just before and just after $T_k$ is committed''.} of $T_k$,
respectively. This can be expressed in the following way by extending
the relation $<_T$ to include local states:
\[T_k\ changes\   x\  from\    \sigma^{i}_{x}\ to\   \sigma^{i+1}_{x}  \iff
(\sigma^{i}_{x} <_T T_k) \wedge (T_k <_T \sigma^{i+1}_{x})\] 
Let   $<^{+}_{T}$ be   the transitive   closure  of the   extended relation
$<_T$. When we 
consider only local states, we get the following {\em happened-before}
relation denoted $<_{LS}$ (which is similar to Lamport's
happened-relation defined on process events \cite{L78} in the
process/message-passing model):
\begin{definition}
\label{precedence between states}
(Precedence on local states,  denoted $<_{LS}$)
\[\sigma^{i}_{x}\   <_{LS}    \sigma^{j}_{y}  \iff   \sigma^{i}_{x}    
 <^{+}_{T}   \sigma^{j}_{y}\] 
\end{definition}

\begin{figure}[htb]
\vspace{0.5cm}
\centering
\input{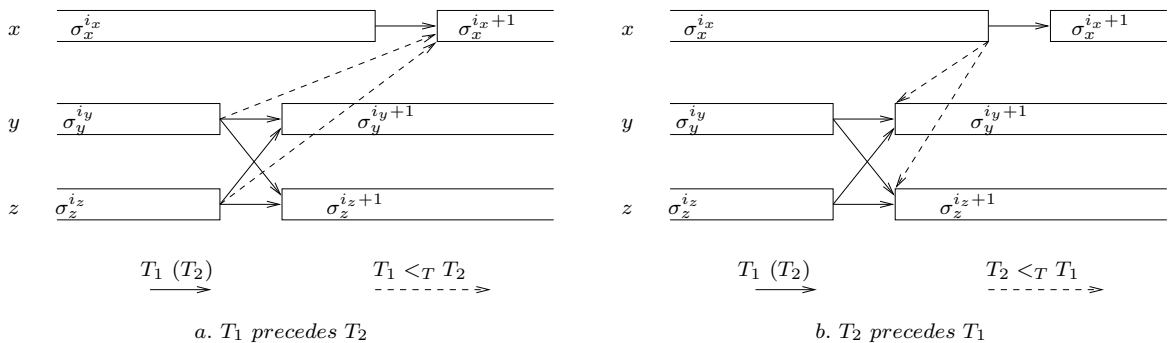}
\caption{Partial Order on Local States}
\label{dipendenza}
\end{figure}

As the relation $<_T$ defined on transactions is a partial order, it
is easy to see that the relation $<_{LS}$ defined on local states is
also a partial order. Figure \ref{dipendenza} shows examples of
relation $<_{LS}$. It considers three data objects $x$, $y$, and $z$,
and two transactions $T_1$ and $T _2$. Transactions are defined in the
following way:

\hspace{5.0cm} $T_1:\ R_1(x); \ W_1(y);\ W_1(z);\ commit_1$ 

\hspace{5.0cm} $T_2:\ R_2(y);\ W_2(x);\ commit_2$

\vspace{0.5cm} 

As there is a read-write conflict on $x$, two serialization orders are
possible. Figure \ref{dipendenza}.a displays the case $T_1 <_T T_2$
while Figure \ref{dipendenza}.b displays the case $T_2 <_T T_1$.  Each
horizontal axis depicts the evolution of the state of a data object.
For example, the second axis is devoted to the evolution of $y$:
$\sigma^{i_y}_y$ and $\sigma^{i_y+1}_y$ are the states of $y$ before
and after $T_1$, respectively.

Let us consider Figure \ref{dipendenza}.a. It shows that $W_1(y)$ and
$W_1(z)$ add four pairs of local states to the relation $<_{LS}$,
namely:
\[\sigma^{i_y}_y<_{LS} \sigma^{i_y+1}_y\]  \[
\sigma^{i_z}_z<_{LS}\sigma^{i_z+1}_z\] \[ 
\sigma^{i_y}_x<_{LS} \sigma^{i_z+1}_z\] \[
\sigma^{i_z}_z<_{LS}\sigma^{i_y+1}_y\] 
Precedence on local states, due to write operations  of
transactions $T_1$ and $T_2$,  are indicated with continuous arrows,
while the ones due to the serialization order are indicated in dashed
arrows\footnote{This shows dependencies between local states can be of
two types. The ones that are due to each transaction taken individually, 
and the ones that are due to conflicting operations issued by 
distinct transactions  (i.e., due to the 
serialization order).}. Figure \ref{dipendenza}.b shows which
precedences are changed when the serialization order is reversed.

%-------------------------------------------------------------------------

\subsection{Consistent Global States}
A {\em global state} of the database is a set of local states, one
from each data object. A global state $G =
\{\sigma^{i_1}_1,\sigma^{i_2}_2, \ldots, \sigma^{i_m}_m\}$ is {\em
  consistent} if it does not contain two dependent local states, i.e.,
if:\[\forall x,y \in [1,\ldots m] \Rightarrow \neg (\sigma^{i_x}_{x}
<_{LS} \sigma^{i_y}_{y})\] Let us consider again Figure
\ref{dipendenza}.a.  The three global states
$(\sigma^{i_x}_{x},\sigma^{i_y}_{y},\sigma^{i_z}_{z})$,
$(\sigma^{i_x}_{x},\sigma^{i_y+1}_{y},\sigma^{i_z+1}_{z})$ and
$(\sigma^{i_x+1}_{x},\sigma^{i_y+1}_{y},\sigma^{i_z+1}_{z})$ are
consistent. The global state
$(\sigma^{i_x+1}_{x},\sigma^{i_y}_{y},\sigma^{i_z+1}_{z})$ is not
consistent either
\remove{\footnote{Of course, only one of these reasons is
  sufficient to claim this global checkpoint is not consistent.}}
because $\sigma^{i_y}_{y} <_{LS} \sigma^{i _x+1}_{x}$ (due to the fact
$T_1 <_T T_2$) or because $\sigma^{i_y}_{y} <_{LS} \sigma^{i_z+1}_{z}$
(due to the fact $T_1$ writes both $y$ and $z$). Intuitively, a
non-consistent global state of the database is a global state that
could not be seen by any omniscient observer of the database.  It is
possible to show that, as in the process/message-passing model, the
set of all the consistent global states is a partial order
\cite{HU95}.

%-------------------------------------------------------------------------
\subsection{Consistent Global Checkpoints} \label{cons}
A {\em local checkpoint} (or equivalently a {\em data checkpoint}) of
a data object $x$ is a local state of $x$ that as been saved in a safe
place\footnote{For example, if $x$ is stored on a disk, a copy is
  saved on  another disk.} by the  data manager of  $x$.  So, all the local
checkpoints are local 
states, but only a subset of local states are defined as local
checkpoints.  Let $C^{i}_{x}$ denote the $i$-th local checkpoint of
$x$; so, $C^{i}_{x}$ corresponds to some $\sigma^{j}_{x}$ with $i \leq
j$. A {\em global checkpoint} is a set of local checkpoints one for
each data object. It is {\em consistent} if it is a consistent global
state.

We assume that all initial local states are checkpointed. Moreover, we
also assume that, when we consider any point of an execution $E$, each
data object will eventually be checkpointed.

%-------------------------------------------------------------------------
%-------------------------------------------------------------------------

\section{Dependence on Data Checkpoints}
\label{DGC}
\subsection{Introductory Example}
As indicated  in  the previous section,  due to  write operations  of each
transaction,  or due to
the serialization order, transactions create dependencies among local
states of data objects. Let us consider the following 7 transactions
accessing data objects $x$, $y$, $z$ and $u$:
\vspace{0.5cm}

\hspace{5.0cm} $T_1:\ R_1(u);\ W_1(u) ;\ commit_1$

\hspace{5.0cm} $T_2:\ R_2(z);\ W_2(z);\ commit_2$

\hspace{5.0cm} $T_3:\ R_3(z);\ W_3(z);\ W_3(x);\ commit_3$

\hspace{5.0cm} $T_4:\ R_4(z);\ R_4(u);\ W_4(z);\ commit_4$

\hspace{5.0cm} $T_5:\ R_5(z); \ W_5(y);\ W_5(z);\ commit_5$

\hspace{5.0cm} $T_6:\ R_6(y);\ W_6(y);\ commit_6$

\hspace{5.0cm} $T_7:\ R_7(x);\ W_7(x);\ commit_7 $

\begin{figure}[t]
\centering
\vspace{0.5cm}
\input{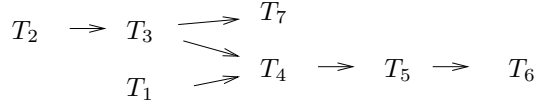}
 \caption{A Serialization Order}
\label{serialization}
\end{figure}

\begin{figure}[t]
\centering
\vspace{0.5cm}
\input{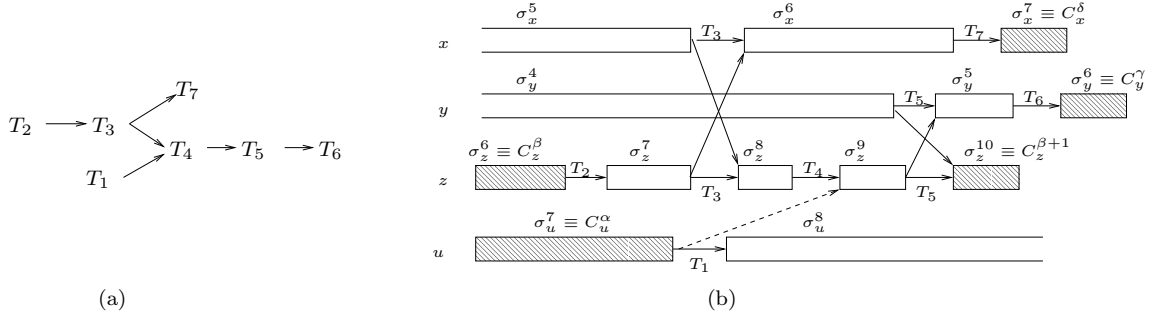}
 \caption{Data Checkpoint Dependencies}
\label{z-path}
\end{figure}

\vspace{0.5cm}
Figure \ref{serialization} depicts the serialization imposed by the
concurrency control mechanism. Figure \ref{z-path} describes
dependencies between local states generated by this execution. Five
local states are defined as data checkpoints (they are indicated by
dark rectangles). We study dependencies between those data checkpoints.
Let us first consider $C^{\alpha}_{u}$ and $C^{\gamma}_{y}$.
$C^{\alpha}_{u}$ is the (checkpointed) state $u$ before $T_1$ wrote
it, while $C^{\gamma}_{y}$ is the (checkpointed) state of $y$ after
$T_6$ wrote it (i.e., just after $T_6$ is committed). The
serialization order (see Figure \ref{serialization}) shows that $T_1
<_T T_6$, and consequently $C^{\alpha}_{u} <_{LS} C^{\gamma}_{y}$,
i.e., the data checkpoint $C^{\gamma}_{y}$ is causally dependent \cite{L78}
on  the data checkpoint $C^{\alpha}_{u}$ (Figure   \ref{z-path} shows that
there is a directed path of local states from  $C^{\alpha}_{u}$ to
 $C^{\gamma}_{y}$). Now let us consider the pair of
data checkpoints consisting of $C^{\alpha}_{u}$ and $C^{\delta}_{x}$.
Figure \ref{z-path} shows that $C^{\alpha}_{u}$ precedes $T_1$, and
that $C^{\delta}_{x}$ follows $T_7$. Figure \ref{serialization}
indicates that $T_1$ and $T_7$ are not connected in the serialization
graph. So, there is no causal dependence between $C^{\alpha}_{u}$ and
$C^{\delta}_{x}$ (Figure     \ref{z-path} shows that  there  is no directed
path from   $C^{\alpha}_{u}$ to $C^{\delta}_{x}$).
But, as the reader can check, there is no consistent
global checkpoint including both $C^{\alpha}_{u}$ and $C^{\delta}_{x}$
(\footnote{Adding $C^{\gamma}_{y}$ and $C^{\beta}_{z}$ to
  $C^{\alpha}_{u}$ and $C^{\delta}_{x}$ cannot produce a consistent
  global state as $C^{\beta}_{z} <_{LS} C^{\delta}_{x}$. Adding
  $C^{\beta+1}_{z}$ instead of $C^{\beta}_{z}$ has the same effect as
  $C^{\alpha}_{u} <_{LS} C^{\beta+1}_{z}$.}). So there is an {\em
  hidden} dependence between $C^{\alpha}_{u}$ and $C^{\delta}_{x}$
which  prevents them to belong to the same consistent global checkpoint.
We now  provide a definition of dependence that takes into account
both causal dependencies and hidden dependencies.

%-------------------------------------------------------------------------

\subsection{Dependence Path}
\label{dp}
\begin{definition}(Interval)
  A checkpoint interval $I^{i}_{x}$ is associated with data checkpoint
  $C^{i}_{x}$. It consists of all the local states $\sigma^{k}_{x}$
  such that:
\[(\sigma^{k}_{x}=C^{i}_{x}) \vee  (C^{i}_{x} <_{LS}  \sigma^{k}_{x} <_{LS}
  C^{i+1}_{x})\] 
\end{definition}

As an example, Figure \ref{z-path} shows that $I^{\beta}_{z}$ includes
4 consecutive local states of $z$. Note that, due to the assumptions
on data checkpoints stated in Section \ref{cons}, any local state
belongs to exactly one interval. Let us call an edge of the partial
order on local states a dependence edge.

\begin{definition}(Dependence  Path)\footnote{This  definition  generalizes
the Z-path notion introduced in \cite{NX95} for 
asynchronous   message-passing   systems.  A   Z-path    is a  sequence  of
messages.  While   a   message is   a ``concrete    entity'',  a dependence
edge     is an ``abstract entity''.  See     Section \ref{distcomp}. So, as
shown by the next theorem,
the  {\it dependence edge} abstraction allows to extend results
of \cite{NX95,W97,BHR} to data checkpoints. }
  
  There is a dependence path ($DP$) from a data checkpoint $C^{i}_{x}$
  to $C^{j}_{y}$ (denoted $C^{i}_{x} \stackrel{DP}{\rightarrow}
  C^{j}_{y}$) iff:

\noindent
i) $x=y$ and $i<j$; or

\noindent
ii) there is a sequence  $(d_1,d_2, \ldots ,d_r)$ of dependence edges, such
that: 

\begin{itemize}

\item[1)]
 $d_1$ starts after $C^{i}_{x}$;
 
\item[2)] $\forall q:\ 1\leq q < r$, $d_q$: let $I^{k}_{z}$ be the
  interval in which $d_q$ arrives; then $d_{q+1}$ starts in the same
  or in a later interval (i.e., an interval $I^{h}_{z}$ such that $k
  \leq h$)\footnote{Note that $d_{q+1}$ can start before $d_{q}$
    arrives. This is where the dependence is ``hidden''. If $\forall q\ 
    d_{q+1}$ starts after $d_q$ arrives, then, the dependence path
    $(d_1,d_2, \ldots ,d_r)$ is purely causal.};

\item[3)]
$d_n$ arrives before $C^{j}_{y}$.
\end{itemize}
\end{definition}

\remove{
This definition generalizes the Z-path notion introduced in \cite{NX95} for
asynchronous   message-passing   systems.  A   Z-path    is a  sequence  of
messages.  So, 
while   a   message is  a ``concrete   entity'',  a dependence  edge  is an
``abstract entity''.    So, the  dependence   edge notion   is definitively
more abstract.  But, while the   message-passing model remains simple 
(there is   a  one-to-one correspondence between  dependence
edges  and messages),  transaction-based  systems are more sophisticated as,
in those systems,  each  transaction  is particular  and generates 
a specific   number   of   dependence  edges  (moreover,  different
transactions a priori generate  distinct numbers of dependency edges). 
(See  also   Section \ref{distcomp}.)  
}
%-------------------------------------------------------------------------
\subsection{Necessary and Sufficient Condition}
\label{cns}

\begin{theorem} \label{consi}
Let ${\cal{I}}\subseteq\{1,\ldots,m\}$ and 
${\cal{S}}=\{C^{i_x}_x\}_{x\in  {\cal{I}}}$ be a set of data
checkpoints.
Then ${\cal{S}}$ is a part of a consistent global 
checkpoint if and only if: 
 $$({\cal P})~~~\forall x,y \in {\cal I}~:~
 \neg(C^{i_x}_x  \stackrel{DP}{\rightarrow} C^{i_y}_{y})$$

\end{theorem}

\begin{proof}

\noindent
{\bf  Sufficiency.} We  prove that  if  $({\cal P})$ is  satisfied  then
${\cal{S}}$  can be included  in a consistent global checkpoint.  Let
us consider the global checkpoint defined as follows:

\begin{itemize}
\item if $x\in {\cal{I}}$, we take $C^{i_x}_x$;
\item if $x\not\in {\cal{I}}$, for each $y\in {\cal{I}}$ we consider
  the integer $m_x(y)={\min}\{i~|~ \neg(C^{i}_x
  \stackrel{DP}{\rightarrow} C^{i_{y}}_{y})\}$ (with $m_x(y)=0$ if
  $i_y=0$ or if this set is empty).  Then we take $C^{i_x}_x$ with
  $i_x=\max_{y\in{\cal{I}}}(m_x(y))$. Let us note that, from that
  definition, it is possible that $i_x=0$ (in that case, $C^{i_x}_x$
  is an initial data checkpoint).
\end{itemize}
By construction, this global checkpoint satisfies the two following
properties :
\begin{equation} \label{eq1}
\forall x \not \in {\cal{I}},~\forall y\in  {\cal{I}}~:~  \neg(C^{i_x}_x  \stackrel{DP}{\rightarrow}
C^{i_{y}}_{y})
\end{equation}
\begin{equation} \label{eq2}
\forall x \not \in {\cal{I}}~{\rm such~that~}i_x>0,~\exists  z\in
{\cal{I}}~:~(i_z>0) \wedge (C^{i_{x}-1}_x
\stackrel{DP}{\rightarrow}  C^{i_{z}}_{z})
\end{equation}

\noindent
We show that $\{C^{i_1}_1, C^{i_2}_2, \ldots,C^{i_m}_m\}$ is
consistent. Assume the contrary. So, there exists $x$ and $y$ and a
dependence edge $d$ that starts after $C^{i_x}_{x}$ and arrives before
$C^{i_y}_{y}$. So, it follows that:

\begin{equation} \label{pippo}
 (i_y>0) \wedge (C^{i_{x}}_x \stackrel{DP}{\rightarrow} C^{i_{y}}_{y})
\end{equation}
\noindent
Four cases have to be considered:

\begin{enumerate}
\item $x\in {\cal{I}}$, $y\in {\cal{I}}$. (3) is contradicted by
  assumption $({\cal P})$.
\item $x\in {\cal{I}}$, $y\not\in {\cal{I}}$. Since $i_y>0$, from
  (\ref{eq2}) we have: $\exists z \in {\cal{I}}~:~(i_z>0) \wedge
  (C^{i_{y}-1}_j \stackrel{DP}{\rightarrow} C^{i_{z}}_{z})$.
  
  As, at data $x$ both the dependence edge ending the path
  $C^{i_x}_{x} \stackrel{DP}{\rightarrow} C^{i_y}_{y}$, and the
  dependence edge starting the path $C^{i_y-1}_{y}
  \stackrel{DP}{\rightarrow} C^{i_z}_{z}$ belong to the same interval,
  we conclude from (\ref{eq2}) that $\exists z \in {\cal I}~:~ (i_z>0)
  \wedge (C^{i_x}_x \stackrel{DP}{\rightarrow} C^{i_{z}}_{z})$ which
  contradicts the assumption $({\cal P})$.
  
\item $x\not\in {\cal{I}}$, $y \in {\cal{I}}$.  (\ref{pippo})
  contradicts (\ref{eq1}).
  
\item $x\not\in {\cal{I}}$, $y \not\in {\cal{I}}$. Since $i_y>0$, from
  (\ref{eq2}) we have: $\exists z\in {\cal{I}}~:~(i_z>0) \wedge
  (C^{i_{y}-1}_y \stackrel{DP}{\rightarrow} C^{i_{z}}_{z})$.
  
  As in case 2, we can conclude that $\exists z \in {\cal I}~:~
  (i_z>0) \wedge (C^{i_x}_x \stackrel{DP}{\rightarrow} C^{i_{z}}_{z})$
  which contradicts (\ref{eq1}).
 \end{enumerate}

\noindent  {\bf Necessity.} We  prove  that,  if there  is a  consistent
global checkpoint $\{C^{i_1}_1, C^{i_2}_2, \ldots,C^{i_n}_n\}$
including ${\cal{S}}$, then property ${\cal P}$ holds for any
${\cal{I}}\subseteq\{1,\ldots,m\}$. Assume the contrary. So, there
exist $x\in {\cal{I}}$ and $y\in {\cal{I}}$ such that $
(C^{i_x}_x \stackrel{DP}{\rightarrow} C^{i_{y}}_{y})$.  From the
definition of $\stackrel{DP}{\rightarrow}$, there exists a sequence of
dependence edges $d_1, d_2, \dots, d_p$ such that:

\begin{tabular}{lll}
& $d_1\ starts\ in\ I_x^{i_x}$, &\\
$d_1\ arrives\ after\ I_{x_1}^{i_1}$, & $d_2\ starts\ in\ I_{x_1}^{j_1}$ &
with $i_1 \leq j_1$\\
& $\ldots$ &\\
$d_{p-1}\ arrives\  in\ I_{x_{p-1}}^{i_{p-1}}$, & $d_p\ starts\ in\ 
I_{x_{p-1}}^{j_{p-1}}$ & with $j_{p-1} \leq i_{p-1}$\\
$d_p\ arrives\ in\ I_y^{i_y-1}$ & &
\end{tabular}

We show by induction on $p$ that, $\forall t \geq i_y$, $C^{i_x}_x$ and
$C^{t}_{y}$ cannot belong  to  the  same consistent global checkpoint.\\
\\
{\em Base step}. $p=1$. In this case, $d_1$ starts after $C^{i_x}_{x}$
and arrives before $C^{i_y}_{y}$, and consequently the pair
($C^{i_x}_{x},C^{i_y}_{y}$) cannot belong to a consistent global
checkpoint.
\\
{\em Induction step}. We suppose the result true for some $p\geq 1$
and show that it holds for $p+1$.  We have:

\begin{tabular}{lll}
& $d_1\ starts\ in\ I_x^{i_x}$, &\\
& $\ldots$ &\\
$d_{p}\ arrives\  in\    I_{x_{p}}^{i_{p}}$,  &  $d_{p+1}\    starts\   in\
I_{x_{p}}^{j_{p}}$ & with $i_{p} \leq j_{p}$\\ 
$d_{p+1}\ arrives\ in\ I_y^{i_y-1}$ & &
\end{tabular}

From the assumption induction applied to the path of dependence edges
$d_1, \ldots, d_p$, we have : for any $t \geq i_p+1$, $C^{i_x}_x$ and
$C^{t}_{x_p}$ cannot belong to the same consistent global checkpoint.
Moreover, $d_{p+1}$ starts in $I_{x_{p}}^{j_{p}}$ and arrives in
$I_y^{i_y-1}$ imply that, for any $h \leq j_p$ and for any $t\geq
i_y$, $C^{h}_{x_p}$ and $C^{t}_{y}$ cannot belong to the same
consistent checkpoint. Since $i_p \leq j_p$, it follows that no
checkpoint of $x_p$ can be included with $C^{i_x}_x$ and $C^{i_y}_{y}$
to form a consistent global checkpoint.
 \end{proof}

\subsection{Database Systems {\it vs} Message-Passing Systems}
\label{distcomp}
\paragraph{Messages {\it vs} transactions.}
An analogous result for message-passing systems has been designed in
\cite{NX95} and generalized in \cite{BHR}. As indicated in Section \ref{dp},
point-to-point message-passing systems are characterized by  the fact  each
message generates exactly one dependence edge between 
two process local checkpoints. 
In database systems, a dependence edge
is due either to a write operation or to the serialization order. As a
transaction may issue several write operations and is serialized in
some order by the concurrency control mechanism, it follows that it may
generate  a lot of  dependence edges between  data checkpoints. For
example, when a transaction writes $\alpha$ data objects, these writes
establish $\alpha^2$ dependence edges and supplementary edges are
added according to the serialization order.

\paragraph{Consistency of a recovery line.} Let us call {\it Recovery
Line}\footnote{Also called $cut$, when adopting the distributed
computing terminology.} a line joining all the data checkpoints of a
global checkpoint. A recovery line is consistent iff the associated
global checkpoint is consistent. Let us remind black and dashed arrows
introduced in the example of Section \ref{localstates}:  a black arrow
denotes a local
checkpoints precedence created by a transaction, while a dashed arrow
denotes a local checkpoints precedence created by the serialization
order. When considering such black and dashed arrows (see Figure
\ref{z-path}), it is possible to show that a recovery line $\cal L$ is
consistent iff:  \begin{itemize} \item No black arrow crosses $\cal L$. 
\item No dashed arrow crosses $\cal L$ from the right of $\cal L$ to the
left of $\cal L$.  \end{itemize}

In a  message-passing system, a  recovery line  (cut) is  consistent iff no
message crosses   it from  the right   to  the  left \cite{CL85}.  Messages
crossing the  recovery  line from left  to   right are  ``in-transit'' with
respect to the recovery line. 
This intuitively shows that,   in a message-passing  system: (1) a  message
corresponds to  a 
``dashed arrow'', and (2) there is no ``black arrow''. So, it appears that
 consistency   of global checkpoints   is a   problem  more
involved   in database  systems than  in  message-passing systems.  

%-------------------------------------------------------------------------
%-------------------------------------------------------------------------
  
\section{Deriving ``Transaction-Induced'' Checkpointing Protocols}
\label{protocols}

\paragraph{Required Properties.}
If we suppose that the set $S$ includes  only a checkpoint $C^{i_x}_x$, the
previous Theorem leads to an interesting corollary $\cal C$: 

\begin{coro} 
$C^{i_x}_x$ belongs to a consistent global checkpoint if and only if 
$\neg (C^{i_x}_x \stackrel{DP}{\rightarrow} C^{i_x}_{x})$. 
\end{coro}
Providing checkpointing protocols
ensuring property  $\cal C$ is interesting for two reasons:\\
- (1) It avoids to waste time in
taking a data checkpoint that will never be used in any consistent
global checkpoint, and \\
- (2) No domino-effect can ever take place as any
data checkpoint belongs to a consistent global
checkpoint\footnote{When, after a  crash, a data  manager  recovers, it can
  restore its 
  last data checkpoint $C$. It follows from $\cal C$ 
  that $C$ belongs to  a consistent global checkpoint. So  the database 
  can be  restarted as  soon as each   data manager  has restored its  data
  checkpoint contained in a  consistent global checkpoint including $C$.
  Note   that,  when compared  to  message-passing  systems, no ``channel
  state'' has  to be restored.}.

~\\

Moreover, let us consider the following property $\cal P$:  
 ``If it exists, the set ${\cal{S}}_n$
formed by the data checkpoints with the same index $n\geq 0$  (one from
each data object), is a consistent global checkpoint''.
In the following we provide two checkpointing protocols:
\begin{itemize}
\item  
The  first   protocol  (${\cal{A}}$)  guarantees   $\cal C$  for all  local
checkpoints,  and guarantees  $\cal P$   for any value of $n$. 
\item The second protocol (${\cal{B}}$) ensures    $\cal C$  only for a subset 
of   local checkpoints, and  $\cal P$ for some particular values of $n$.
\end{itemize}
Actually,  those   protocols    can   be   seen  as  adaptations to     the
data-object/transactions  model,   of    protocols     developed  for  the
process/message-passing   model.  More  precisely,    protocol  ${\cal{A}}$
corresponds with  Briatico {\it et al.}'s protocol \cite{BCS84}, 
 while     protocol ${\cal{B}}$   corresponds  with   Wang-Fuchs's protocol
\cite{WA93}.

\paragraph{Local Control Variables.}
In both protocols we assume each data manager $DM_x$ has an index
$i_x$, which indicates the index (rank) of the last checkpoint of $x$
(it is initialized to zero).  Moreover, each data manager can take
checkpoint independently ({\em basic checkpoints}), for example, by
using a periodic algorithm which could be implemented by associating a
timer with  each data manager.  A local  timer is set whenever a checkpoint is
taken. When a local  timer expires, a basic checkpoint is taken by the data
manager. Data managers are  directed  to take additional data   checkpoints
({\em forced  checkpoints})  in order to ensure  $\cal C$ or $\cal P$. 
 The decision to take forced
checkpoints is based on the control information piggybacked by
transactions.

A protocol consists of two interacting parts. The first part, shared by  
both  algorithms,   specifies  the   checkpointing-related   actions  of 
transaction managers. The second part defines the  rules  data managers have
to follow to take data checkpoints.

\paragraph{Protocols  ${\cal{A}}$ and ${\cal{B}}$: 
Behavior of a Transaction  Manager.} 
Let $W_{T_i}$ be the write set of a transaction $T_i$ managed by a
transaction manager $TM_i$. We assume each time an operation of $T_i$
is issued by $TM_i$ to a data manager $DM_x$, it returns the value of
$x$ plus its index $i_x$.  $TM_i$  stores in $M_{T_i}$ the maximum value
among the indices of the data objects read or written  by $T_i$.  When
 transaction $T_i$   is committed, the transaction  manager  $TM_i$ sends a
{\sc commit} 
message to each data manager $DM_x$ involved in $W_{T_i}$. Such {\sc commit}
messages piggyback $M_{T_i}$.

\paragraph{Protocol ${\cal{A}}$: Behavior of a Data Manager.}

As far  as checkpointing  is  concerned, the behavior  of  a  data manager
$DM_x$ is   defined  by   the    two following  procedures    namely   {\tt
take-basic-ckpt}  and  {\tt   take-forced-ckpt}. They   defined   the rules
associated with checkpointing.

\begin{description} 
  \item[{\tt take-basic-ckpt(${\cal{A}}$)}]:\\
\hspace*{1cm}   {\bf When}  the timer expires:\\
\hspace*{2cm}   (AB1) $i_x  \leftarrow i_x +1$;\\
\hspace*{2cm}   (AB2)  Take  checkpoint $C_{x}^{i_x}$;\\
\hspace*{2cm}   (AB3) Reset the local timer. 
  
\item[{\tt take-forced-ckpt(${\cal{A}}$)}]: \\
\hspace*{1cm} {\bf  When} $DM_x$ {\bf   receives} {\sc   commit}($M_{T_i}$)
{\bf from} $TM_i$: \\
\hspace*{2cm}  {\bf if} $i_x < M_{T_i}$ {\bf then}\\
\hspace*{3cm}   (A1)  $i_x  \leftarrow  M_{T_i}$;\\
 \hspace*{3cm}   (A2) Take  a (forced) checkpoint $C_{x}^{i_x}$;\\
\hspace*{3cm}    (A3) Reset the local timer. \\
\hspace*{2cm} {\bf endif};\\
\hspace*{2cm}    (A4) process the {\sc  commit} message.
\end{description}

From the increase of the index $i_x$ of a data object $x$, and from 
the rule {\tt take-forced-ckpt(${\cal{A}}$)} (which forces a data checkpoint
whenever $i_x < M_{T_i}$), the condition $\neg (C^{i_x}_x
\stackrel{DP}{\rightarrow} C^{i_x}_{x})$ follows for any
data checkpoint. Actually, this simple protocol ensures that, if
 $C^{i_x}_x   \stackrel{DP}{\rightarrow}   C^{i_y}_{y}$,  then the  index
$i_x$ associated with $C^{i_x}_x$  is strictly lesser than the  index  $i_y$
associated with  $C^{i_y}_y$.   

It follows from the previous  observation that 
if two data checkpoints have the same index, then they cannot be related by
 $\stackrel{DP}{\rightarrow}$.  So, all  the 
 sets ${\cal{S}}_n$ that exist are consistent.  
Note that the      {\tt take-forced-ckpt(${\cal{A}}$)} rule may
produce  gaps in the sequence of indices assigned to data checkpoints
of a  data object $x$. So, from a practical point of view, the 
following remark is interesting:  when  
no data checkpoint of a data object $x$ is indexed by a given value 
$n$, then the  first
data checkpoint of $x$ whose index is greater than $n$, can be included 
in a set containing data  checkpoints indexed by $n$,  to form a  consistent
global checkpoint.  

\paragraph{Protocol ${\cal{B}}$: Behavior of a  Data Manager.}
This protocol introduces a system parameter $Z\geq 1$ known by all
the data managers \cite{WA93}. Only for  subset of data checkpoints whose
index is equal to $a \times Z$ (where $a\geq 0$ is an integer), we have:  $\neg
(C^{aZ}_x  \stackrel{DP}{\rightarrow}  C^{aZ}_{x})$.   Moreover, 
when, $\forall x$,   $C^{aZ}_{x}$ exists, then the     global checkpoint
 ${\cal{S}}_{aZ}$ exists and is consistent. 

 The rule {\tt  take-basic-ckpt(${\cal{B}}$)} is the same to the one of the
protocol  ${\cal{A}}$. In addition to the previous control variables, 
each  data  manager $DM_x$ has    an additional variable  $V_x$,   which is
incremented by $Z$ each time a data checkpoint indexed $aZ$   is  taken.
The rule  {\tt take-forced-ckpt(${\cal{B}}$)} is the following.

\begin{description} 
\item[{\tt take-forced-ckpt(${\cal{B}}$)}]: \\
\hspace*{1cm} {\bf When} $DM_x$ {\bf receives}  {\sc commit}($M_{T_i}$) 
{\bf from}  $TM_i$: \\
\hspace*{2cm}   {\bf if} $V_x < M_{T_i}$ {\bf then} \\
\hspace*{3cm}    (B1)  $i_x \leftarrow \lfloor  M_{T_i}/Z \rfloor \times Z$;\\
\hspace*{3cm}   (B2) Take a (forced)  checkpoint $C_{x}^{i_x}$;\\
\hspace*{3cm}   (B3) Reset the local timer;\\
\hspace*{3cm}    (B4) $V_x \leftarrow V_x + Z$. \\
\hspace*{2cm}   {\bf endif};\\
\hspace*{2cm}    (B5) Process the {\sc commit} message. 

\end{description}

\paragraph{About coordination.}
Compared to previous checkpointing protocols appeared in the
literature \cite{PK92,SON}, which use an explicit coordination among data
managers to get consistent global checkpoints, the proposed  protocols
provide  the same result by using a {\em lazy} coordination which
is  propagated  among  data  managers  by  transactions  (with {\sc commit}
messages). In particular, 
protocol ${\cal{A}}$ starts a ``transaction-induced'' coordination
each time a basic checkpoint is taken; while protocol ${\cal{B}}$
starts a coordination each time a basic checkpoint, whose index is a
multiple of the parameter $Z$, is taken. The latter protocol seems to
be particularly interesting for database systems as it shows a
tradeoff, mastered by a system parameter $Z$, between the number of
forced checkpoints and the extent of rollback during a recovery phase.
The greater $Z$ is, the larger will be the rollback distance.

%------------------------------------------------------------------------
%------------------------------------------------------------------------

\section{Conclusion}
This paper has  presented a formal approach for consistent data
checkpoints in database systems. Given an arbitrary set of data
checkpoints  (including at least a single data checkpoint from a data
manager, and at most a
data checkpoint from each data manager), we answered the following
important question ``Can these data checkpoints be members of a same
consistent global checkpoint?'' by providing a necessary and sufficient
condition.  We have also derived two {\em non-intrusive} data
checkpointing protocols from this condition; these checkpointing protocols
use transactions as a means to diffuse information among data managers.

This paper can also be  seen as a bridge between the area of
distributed computing  and   the area of databases.  We have shown that the
checkpointing problem is harder in 
data-object/transaction systems than in
process/message-passing systems.  From a distributed computing point of
view, we could say that database systems are difficult because they
merge the ``synchronous world'' (every transaction taken individually
has to be perceived as {\it atomic}: it can be seen as a multi-rendezvous among
the objects it is on) and the ``asynchronous world'' (due to relations
among transactions managed by the concurrency control mechanism).

\remove{
By  replacing strictness  with  rigorousness   \cite{BGRS91} in the   model
presented in Section \ref{model} the presented approach  can be extended to
deal with hierarchical combination of transaction management systems (e.g.,
interoperability among heterogeneous databases).
}

\end{document}